\documentclass[[oneside,pdflatex,sn-mathphys-num]{sn-jnl}% Math and Physical Sciences Numbered Reference Style 
\makeatletter
\@twosidefalse  % 关闭双面模式
\@mparswitchfalse  % 关闭页边注切换
\makeatother

\usepackage{graphicx}%
\usepackage{multirow}%
\usepackage{amsmath,amssymb,amsfonts}%
\usepackage{amsthm}%
\usepackage{tabularx}
\usepackage{mathrsfs}%
\usepackage[title]{appendix}%
\usepackage{xcolor}%
\usepackage{textcomp}%
\usepackage{manyfoot}%
\usepackage{booktabs}%
\usepackage{algorithm}%
\usepackage{algorithmicx}%
\usepackage{algpseudocode}%
\usepackage{listings}%
\usepackage{comment}
\usepackage{subfig} % 
\usepackage[numbers]{natbib}
\theoremstyle{thmstyleone}%
\theoremstyle{thmstyletwo}%

\theoremstyle{thmstylethree}%
\normalsize%
\begin{document}
 \large

\title[Article Title]{Structural and Electronic Evolution of Bilayer Nickelates Under Biaxial Strain}

\author[1]{\fnm{H C Regan B.}
\sur{Bhatta}}%\email{hbhatta@ufl.edu}
\equalcont{These authors contributed equally to this work.}

\author[1]{\fnm{Xiaoliang}
\sur{Zhang}}%\email{xiaolian.zhang@ufl.edu}
\equalcont{These authors contributed equally to this work.}

\author*[2]{\fnm{Yong}
\sur{Zhong}}%\email{ylzhong@stanford.edu}
% \affiliation{Stanford Institute for Materials and Energy Sciences, SLAC National Accelerator Laboratory, Menlo Park, CA 94025, USA}

\author*[1]{\fnm{Chunjing}
\sur{Jia}}
\email{ylzhong@stanford.edu, chunjing@ufl.phys.edu}
% \affiliation{Department of Physics, University of Florida, Gainesville, FL 32611, USA}
\affil*[1]{\orgdiv{Department of Physics}, \orgname{University of Florida}, \orgaddress{ \city{Gainesville}, \postcode{32611}, \state{FL}, \country{USA}}}
% \date{\today}
%\maketitle
\affil[2]{\orgdiv{Stanford Institute for Materials and Energy Sciences}, \orgname{SLAC National Accelerator Laboratory}, \orgaddress{\city{Menlo Park}, \postcode{94025}, \state{CA}, \country{USA}}}

\abstract{\large
The discovery of high-{\it T}$_c$ superconductivity around 80K in bilayer nickelate $\text{La}_3\text{Ni}_2\text{O}_7$ under high pressure has expanded the family of high-{\it T}$_c$ superconductors above the nitrogen boiling temperature. Recent studies  have further shown that ambient pressure superconductivity with a {\it T}$_c$ exceeding 40K can be achieved in compressively strained $\text{La}_3\text{Ni}_2\text{O}_7$ thin films, offering a tunable platform for investigating the pairing mechanism in high-{\it T}$_c$ nickelates. A comprehensive understanding of the structural and electronic properties of bilayer nickelate under epitaxial strain is essential to advance this active field. In this work, we employ first-principles calculations to systematically explore the entire rare-earth (Re) series of bilayer nickelates $\text{Re}_3\text{Ni}_2\text{O}_7$ in the realistic orthorhombic {\it Amam} phase under various compressive and tensile strain conditions. We highlight the materials properties change when strain is applied, and compare these results with those observed under high pressure. Our findings show that 2.5\% compressive strain increases the apical Ni-O-Ni bond angle toward 180 degree, and causes the Ni $d_{z^2}$ bands to move away from the Fermi level.  The tight-binding parameters for the 2.5\% compressively strained $\text{La}_3\text{Ni}_2\text{O}_7$ are quite similar to those of the unstrained material, except that the on-site energy difference between the Ni $d_{z^2}$ and $d_{x^2-y^2}$ orbitals increases by about 50 percent. Notably, the absence of the $d_{z^2}$ bands at the Fermi energy under compressive strain contrasts sharply with the electronic structure in the high-pressure {\it Fmmm} phase, suggesting that the presence of $d_{z^2}$ bands at the Fermi energy may not be a requisite for superconductivity. Our study provides critical insights into recent reports of ambient pressure superconductivity in $\text{La}_3\text{Ni}_2\text{O}_7$ and La$_2$PrNi$_2$O$_7$ thin films, offering valuable guidance for optimizing the superconducting properties of strained $\text{Re}_3\text{Ni}_2\text{O}_7$ thin films across the lanthanide series.

}


%\keywords{keyword1, Keyword2, Keyword3, Keyword4}
\maketitle

\newpage
\section*{Introduction}
% The microscopic understanding of the pairing mechanism in high-{\it T}$_c$ cuprates remains an open question even after nearly four decades\cite{bednorz1986,RevModPhys.66.763}. One promising approach to addressing this challenge is to conduct comparative studies with other superconducting materials that possess  similar crystal structures and magnetic properties to cuprates, which can provide valuable insights to eludicate the enigma of high-{\it T}$_c$ superconductivity. The discovery of superconductivity in infinite-layer nickelate thin films has inspired many effects to search for more nickel-based superconductors\cite{li2019,PhysRevX.11.011050,PhysRevB.101.060504,PhysRevB.106.115150,PhysRevLett.124.207004,PhysRevLett.125.147003}. Recently, the signature of superconducting transition around 80 K was first reported in the bilayer nickelate La$_3$Ni$_2$O$_7$ under high pressure\cite{80KSuper}. This finding was further confirmed by the observations of zero resistance and Meissner effect in the La$_2$PrNi$_2$O$_{7}$ compound\cite{wang2024nature}, making nickelates join the club of high-{\it T}$_c$ superconductors above nitrogen boiling temperature. 

The microscopic pairing mechanism in high-$T_c$ cuprates remains an open question despite nearly four decades of extensive research~\cite{bednorz1986,RevModPhys.66.763}. A promising strategy to unravel this long-standing mystery is to conduct comparative studies with other superconducting materials that share similar crystal structures and magnetic properties with cuprates. Such investigations can provide crucial insights into the underlying physics of high-$T_c$ superconductivity. In this context, the discovery of superconductivity in infinite-layer nickelate thin films has sparked significant interest, leading to extensive efforts to explore additional nickel-based superconductors~\cite{li2019,PhysRevX.11.011050,PhysRevB.101.060504,PhysRevB.106.115150,PhysRevLett.124.207004,PhysRevLett.125.147003}. Recently, a superconducting transition at approximately 80~K was reported in bilayer nickelate $\text{La}_3\text{Ni}_2\text{O}_7$ under high pressure~\cite{80KSuper}. This breakthrough was further corroborated by the observation of zero resistance and the Meissner effect in the related compound $\text{La}_2\text{Pr}\text{Ni}_2\text{O}_7$~\cite{wang2024nature}, firmly establishing nickelates as members of the high-$T_c$ superconductor family with transition temperatures exceeding the boiling point of liquid nitrogen.

High pressure plays a crucial role in the realization of superconductivity in bilayer nickelates, which is accompanied by a structural transition from orthorhombic symmetry ({\it Amam}) under ambient pressure to orthorhombic symmetry ({\it Fmmm}) or tetragonal symmetry ({\it I}4/{\it mmm}) under high pressure\cite{80KSuper,Wang2024ACS, wang2024nature, Geisler2024,Luo2024,PhysRevLett.131.126001,PhysRevLett.132.106002,Lu2024,PhysRevB.109.045154}. However, the high pressure conditions are very challenging for many characterization tools, hindering the comprehensive studies of the microscopic mechanism in high-$T_c$  nickelates. The epitaxial technique offers a unique opportunity to overcome the above obstacles through tuning the in-plane lattice parameters to simulate the effects of high pressure. In perovskite materials, it has been well established that epitaxial strain and pressure can have analogous effects on the bond angles and bond lengths of the octahedral units, as seen in systems such as manganites, rare-earth nickelates and transition metal halides\cite{ke2021, steele2019,RANNO2002170,KONISHI1998158,PhysRevB.88.195108,Guo2020,10.1063/1.126248,annurev:/content/journals/10.1146/annurev-matsci-070115-032057}. Unlike high pressure technique that requires sophisticated instrumentation such as diamond anvil cells, strain engineering is a more versatile and accessible approach in experiments. 

The recent studies have demonstrated ambient pressure superconductivity with a onset $T_c$ ($T_{c,\text{onset}}$) of up to 42 K in compressively strained La$_3$Ni$_2$O$_7$ thin films\cite{Ko2024}. Furthermore, La$_2$PrNi$_2$O$_7$ thin films exhibit enhanced stabilization of the Ruddlesden-Popper (R-P) phase, leading to increases in both $T_{c,\text{onset}}$ and $T_{c,\text{zero}}$\cite{liu2025superconductivitynormalstatetransportcompressively,Zhuoyu}. These findings highlight the need for theoretical investigations into the strain-induced structural and electronic properties of bilayer nickelate thin films. DFT calculation without $U$ correction on the biaxial in-plane strain on {\it Amam} La$_3$Ni$_2$O$_7$ shows that with compressive strain, the apical Ni-O-Ni does turn towards 180$^\circ$, mimicing the structural transition towards {\it Fmmm} under high pressure\cite{rhodes2024structural, Zhao2024,Wang2024}. However, several key questions are still waiting to be addressed, including the evolution of the structure and electronic properties of bilayer nickelates with different lanthanide elements in the realistic low-symmetry {\it Amam} phase under biaxial strain, and the similarity or difference of these properties and microscopic band parameters when tuned by strain versus those tuned by pressure.                  

%The structural and electronic modifications of nickelates under strain and pressure have garnered significant attention, particularly in the context of enabling in bilayer nickelates, which exhibit the highest T$_c$ among nickelate superconductors \cite{hou2023work1, zhang2023work2, zhangAmamFmmm, ZhangAmamFmmm2, luo2023work3, lechermann2023work4, christiansson2023work5, lu2024work6, qu2024work7, sakakibara2024work8s+-, yang2023work9s+-, wang2024work10, hou2023work11, yang2023interwork12, lechermann2023work13, shen2023effectivework14, tian2024work15s+-, wu2024superexchangework16, liao2023electronwork17}. DFT calculation without $U$ correction on the biaxial in-plane strain on {\it Amam} La$_3$Ni$_2$O$_7$ shows that with compressive strain, the apical Ni-O-Ni does turn towards 180$^\circ$, mimicing the structural transition towards {\it Fmmm} under high pressure.\cite{rhodes2024structural, Zhao2024} However, a systematic theoretical investigation on the biaxial tuning of structural transition in bilayer nickelates with different lanthanide elements, and how the electronic band structures and the microsocopic tight-binding model paramteres are modified with strain at the realistic case for low symmetry orthorhombic {\it Amam} phase need to address. Key questions include whether strain affects the structure similarly or differently compared to pressure and how these modifications influence the electronic structure and microscopic models relevant to superconductivity.  

To answer these questions, we theoretically investigate strain as an approach to engineer bilayer nickelates at ambient pressure. %, investigating whether strain-induced tuning of the structure and electronic properties can achieve effects similar to those observed under pressure. 
We calculated the structural and electronic properties of $\text{Re}_3\text{Ni}_2\text{O}_7$ in orthorhombic {\it Amam} phase under both compressive and tensile biaxial in-plane strains using first-principles calculations based on density functional theory (DFT). Our DFT calculations with $U$ correction show that a 2.5\% compressive strain causes apical Ni-O-Ni bond to tilt towards $180^\circ$, similar to the effect of 6 GPa pressure, although with strain-tuning they do not fully reach this angle. Notably, the band structure near the Fermi level exhibits different trends compared with high pressure. Unlike the high pressure scenario in the {\it Fmmm} phase, where the $d_{z^2}$ bands become more dispersive and the bonding $d_{z^2}$ band moves above the Fermi level, compressive strain results in the flattening of the $d_{z^2}$ bands, with the bonding $d_{z^2}$ band shifts to lower energy, moving further away from the Fermi level. 
%Previous studies employed a two-Ni-atom, four-orbital tight-binding model to investigate the band structure and the electron and hole pockets on the Fermi surface in the \textit{Fmmm} phase \cite{PhysRevLett.131.126001,Zhang2024}. However, our analysis indicates that strained systems do not transition to the \textit{Fmmm} phase but instead exhibit reduced symmetry. To address this, we constructed a four-Ni-atom, eight-orbital tight-binding model to examine the effects of strain on the hopping parameters. Using the post-DFT package Wannier90 \cite{Pizzi2020}, we derived accurate hopping parameters and site energies for both the unstrained case and the 2.5\% compressive strain case.  
Analysis of the tight-binding  parameters reveals that the observed discrepancies in the band structure comes alongside with the large compressive-strain-induced increase of the on-site energy difference between Ni $d_{x^2-y^2}$ and $d_{z^2}$ by more than 50\%. Additionally, 2.5\% compressive strain increases the inter-layer hopping of Ni $d_{z^2}$ and intra-layer hoppings of Ni $d_{x^2-y^2}$. This microscopic understanding of strain-tuning provides valuable insights into how strain influences the electronic structure of the system and its implications for the mechanism underlying unconventional superconductivity in nickelates.

%which increases from 0.74 to 1.20 eV when 2.5\% compressive in-plane strain is applied, making the electron occupation on Ni $d_{z^2}$ increased upon compressive strain. Upon 2.5\% compressive strain, the nearest-neighbor intra-layer hopping for Ni $d_{z^2}$ decreases, while the inter-layer hopping for Ni $d_{z^2}$ orbitals via the apical oxygen increases. Additionally, the nearest-neighbor intra-layer hopping for Ni $d_{x^2-y^2}$ increases. This microscopic understanding of strain-tuning provides valuable insights into how strain influences the electronic structure of the system and its implications for the mechanism underlying unconventional superconductivity in nickelates.

\begin{figure*}[htp!]
  \centering  \includegraphics[width=\linewidth]{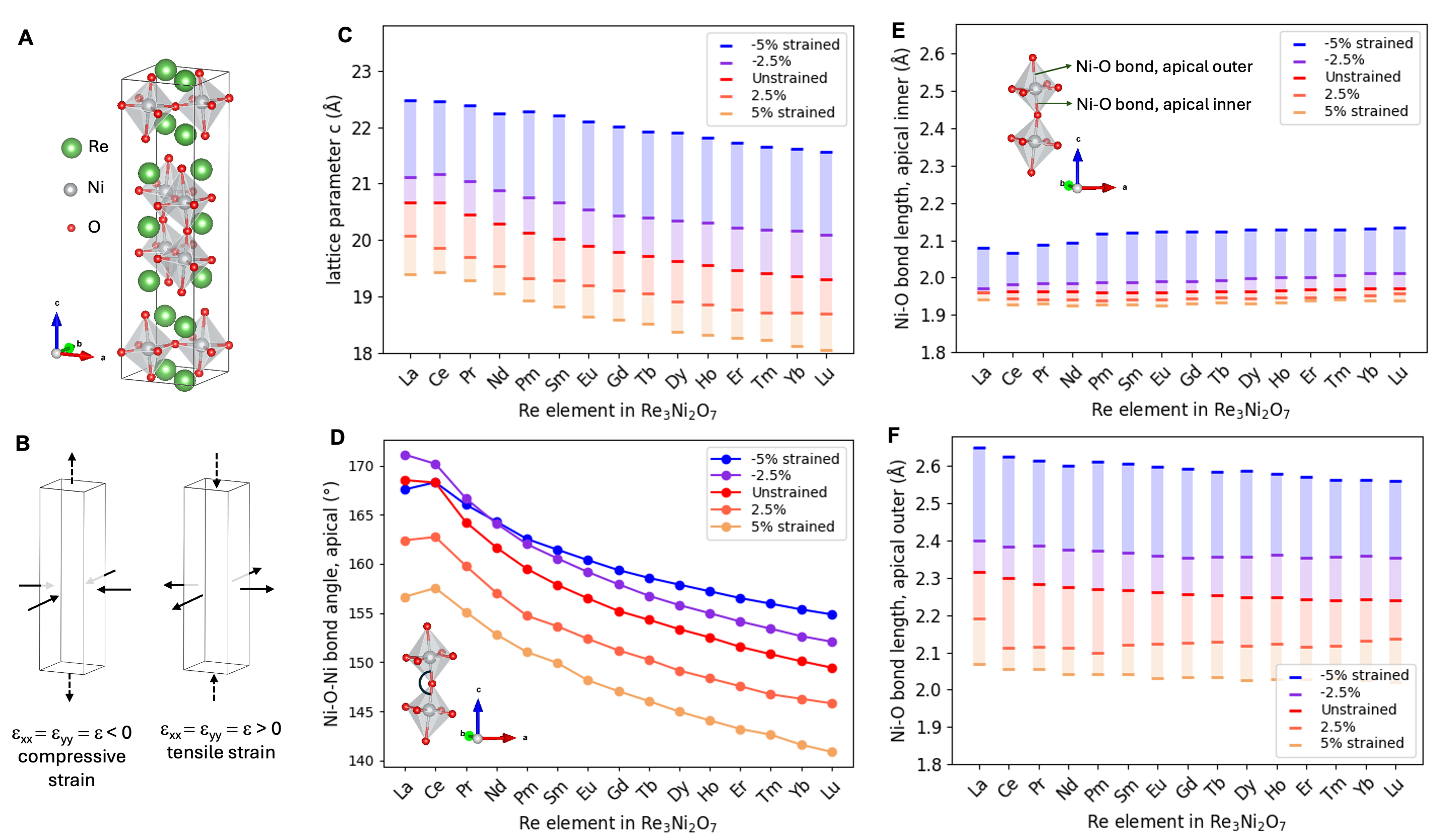}
  \caption{\textbf{Strain-tuned bilayer nickelates Re$_3$Ni$_2$O$_7$: Effects on lattice parameters, bond angles, and bond lengths across different rare-earth elements (Re) using first-principles calculations. A}, Crystal structure of bilayer nickelates $\text{Re}_3$$\text{Ni}_2$$\text{O}_7$. \textbf{B}, Schematic illustration of lattice parameter changes under biaxial in-plane compressive and tensile strain. \textbf{C}, Lattice parameter $c$ variations upon different strains.  \textbf{D},  Apical octahedral angle (Ni-O-Ni) changes with applied strain. The inset shows a visualization of the apical octahedral angle. \textbf{E}, Apical Ni-O bond length at the inner of the bilayer under different strains. The inset distinguishes the two types of apical Ni-O bonds: inner and outer ones. \textbf{F},  Apical Ni-O bond length at the outer of the bilayer under different strains. }
  \label{fig:Fig1}
\end{figure*}

%We utilized Wannier downfolding \cite{PhysRevB.65.035109,Pizzi2020} to extract the hopping parameters for a sixteen-orbital model, taking Ni \(3d_{3z^2-r^2}\) and \(3d_{x^2-y^2}\) Wannier orbitals. This procedure was implemented using Wannier90, interfaced with Quantum Espresso for the band structure calculations. The Ni 3$d_{x^2-y^2}$ bands are entangled with bands with dominant La orbital content, due to energy overlap above 1eV. To isolate the target bands associated with the Ni 3$d_{x^2-y^2}$ and 3$d_{z^2}$ orbitals, we employed the disentanglement procedure \cite{PhysRevB.65.035109} implemented in Wannier90, ensuring a more accurate representation of the low-energy tight-binding description.

\section*{Results}

\subsection*{Strain-induced Structural Change}

The crystal structure of bilayer nickelates at ambient pressure in the orthorhombic {\it Amam} phase is shown in Fig. \ref{fig:Fig1}\textbf{A}, where Re represents rare-earth elements such as La, Nd, and Pr. Biaxial in-plane compressive or tensile strain with $\epsilon_{xx} = \epsilon_{yy} = \epsilon$ is applied, with the $c$-lattice parameter and atomic positions allowed to relax freely, as depicted in Fig. \ref{fig:Fig1}\textbf{B}. DFT+$U$ calculations with $U = 3.5$eV were implemented for structural relaxation. This value of $U$ was previously demonstrated to accurately reproduce the band structure of ARPES measured La$_3$Ni$_2$O$_7$\cite{UValueyang2024orbital}, and consistent with other first-principles studies~\cite{Geisler2024,PhysRevB.108.165141,geislerUvalue}. Structural relaxation calculations reveal that biaxial in-plane compressive strain increases the lattice parameter $c$, while biaxial in-plane tensile strain decreases it, as expected across the entire lanthanide series illustrated in Fig. \ref{fig:Fig1}\textbf{C}. This increase in $c$ is accompanied by an increase in the apical bond angle under 2.5\% compressive strain, as illustrated in Fig. \ref{fig:Fig1}\textbf{D}. Specifically, the apical Ni–O–Ni bond angle for La$_3$Ni$_2$O$_7$ increases from 168.5$^\circ$ to 171.1$^\circ$ under biaxial in-plane 2.5\% compressive strain.  Compressive strain at 2.5\% increases the apical bond angle for all materials across the lanthanide series; however, the strain-induced apical bond angle increase has an upper limit: it cannot drive the apical angle to 180$^\circ$ or induce a structural transition to orthorhombic $Fmmm$ or tetragonal {\it I}4/${mmm}$ phase. For La and Pr, further increasing the biaxial in-plane compressive strain up to 5\% reduces the apical bond angle, moving it away from 180$^\circ$. Across the lanthanide series, the apical Ni-O-Ni angle in the unstrained structure decreases when the lanthanide element gets heavier. For elements heavier than Nd, compressive strain monotonically increases the apical Ni-O-Ni angle, while tensile strain monotonically decreases the apical Ni-O-Ni angle, in the range of our study.

\begin{figure*}[hptb] 
    \centering \includegraphics[width=\columnwidth]{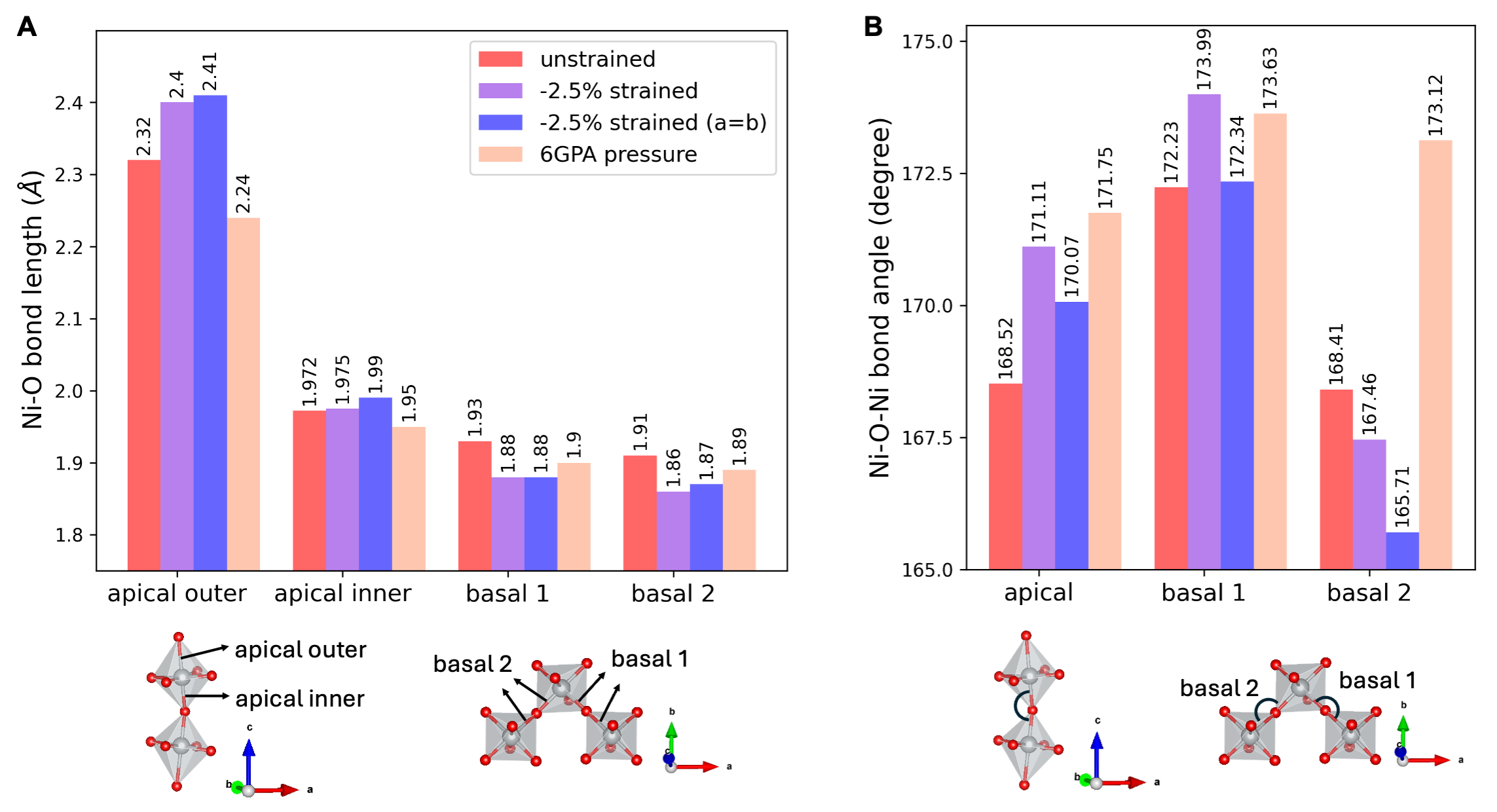}  % Adjust width as needed
\caption{\textbf{Comparison of the effects of compressive strain and pressure on the bilayer La$_3$Ni$_2$O$_7$ structure}. \textbf{A}, Ni-O bond lengths and \textbf{B} the Ni–O–Ni bond angles are shown for the unstrained structure, isotropic compressive strain of 2.5\%, anisotropic compressive strain of 2.5\% (averaged in-plane strain with equal lattice parameters for a and b), and under 6 GPa pressure. Structural relaxations calculations are implemented with DFT+$U$ with $U=3.5$eV.}
    \label{bar}
\end{figure*}

We further analyzed the basal Ni–O–Ni bond angles under biaxial in-plane compressive and tensile strain, as shown in Fig.~S1 of the Supplementary Materials. For La, Pr, and Nd under 2.5\% compressive strain, the two basal angles deviate further from each other compared to the unstrained case: the larger basal Ni–O–Ni angle increases, while the smaller angle decreases. This indicates that biaxial in-plane compressive strain enhances in-plane anisotropy along the basal 1 and basal 2 directions—the two in-plane directions connecting nearest-neighbor Ni atoms—in bilayer nickelates for La, Pr, and Nd.

The apical Ni–O bond lengths, as shown in Fig.~\ref{fig:Fig1}\textbf{E} and \textbf{F}, exhibit minimal change in the apical inner bonds under strain, including 2.5\% compressive strain and all tensile strain cases. In contrast, the apical outer Ni–O bond length shows a pronounced response to strain. This suggests that the two Ni–O layers maintain a relatively stable separation under strain, with the primary variation in lattice parameters along the $c$-axis occurring at the outer Ni–O bond length and other regions outside of the Ni-O bilayers. For 5\% compressive strain, the inner apical Ni–O bond lengths begin to increase significantly across all materials, indicating that the two Ni–O layers start to move apart as high compressive in-plane strain is applied.

We compared the structural properties under biaxial in-plane 2.5\% compressive strain, with 6GPa pressure, and biaxial in-plane compressive strain applied anisotropically along $a$ and $b$ for an average strain of -2.5\% with lattice parameters satisfying $a=b$, with the results shown in Fig. 2. In the latter case, we simulated the conditions of epitaxial strain where bilayer nickelates are grown on substrates with in-plane lattice parameters $a=b$, which are 2.5\% smaller than the average in-plane lattice parameters of the bilayer nickelates. Our DFT+$U$ calculations show that the behaviors of Ni-O bond lengths under isotropic -2.5\% strain match that observed under an average -2.5\% strain with
$a=b$, and Ni–O–Ni bond angles under isotropic -2.5\% strain are all larger than those at the an average -2.5\% strain with
$a=b$. When comparing these trends to the structural properties at 6 GPa, we find that -2.5\% strain partially resembles the effects of pressure. Specifically, under -2.5\% strain, the apical bond angle is comparable to but smaller than that at 6 GPa, while the basal bond angles are significantly more asymmetric. Additionally, the inner apical Ni–O bond length is similar to that at 6 GPA, but the outer apical Ni–O bond length differs substantially.

\begin{figure*}[ht!]
  \centering
  \includegraphics[width=\linewidth]{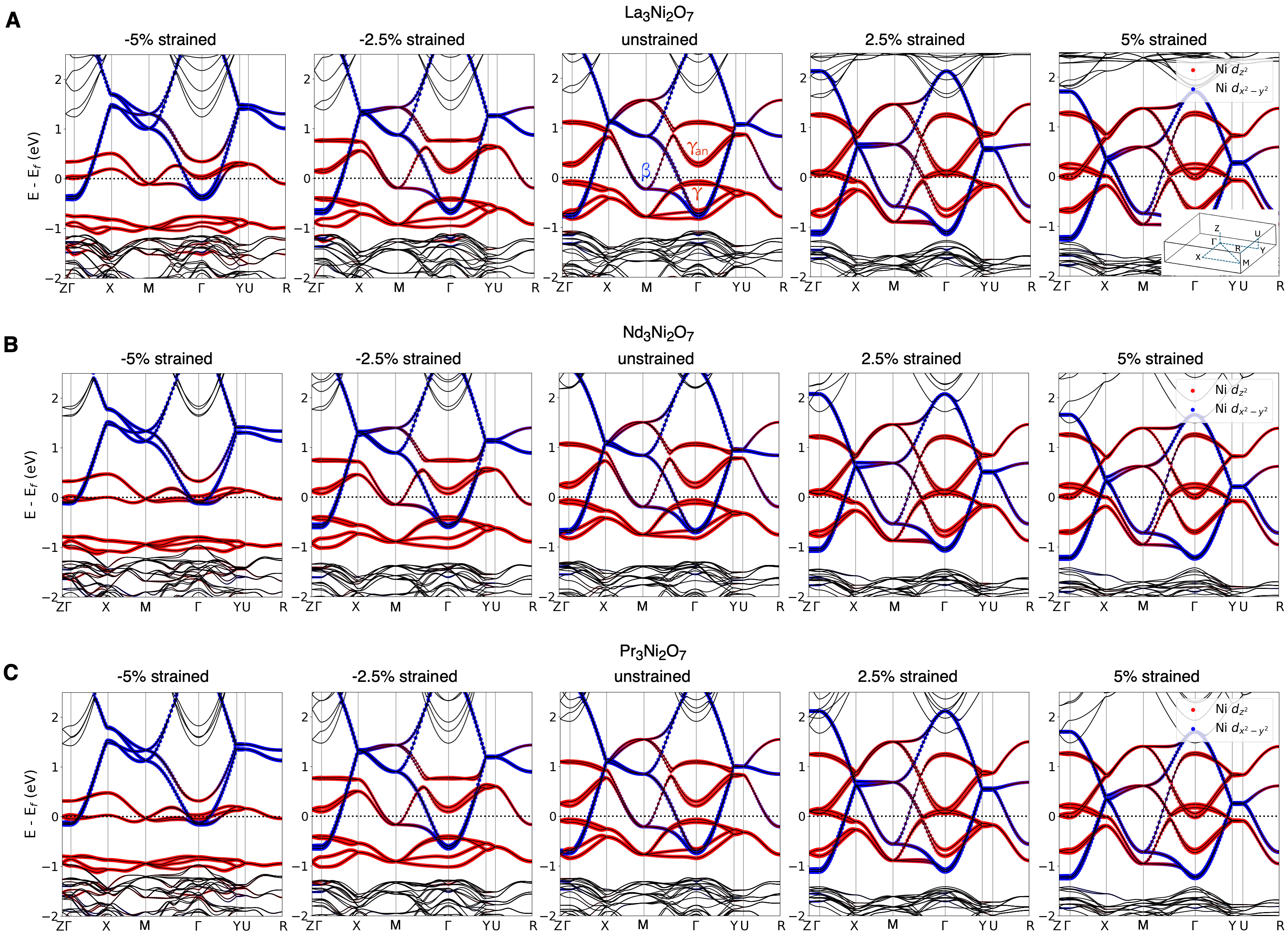}
\caption{\textbf{Orbital-resolved band structure of bilayer nickelates under different in-plane strains.}  
\textbf{A} $\text{La}_3\text{Ni}_2\text{O}_7$, \textbf{B}
$\text{Nd}_3\text{Ni}_2\text{O}_7$, and \textbf{C}
$\text{Pr}_3\text{Ni}_2\text{O}_7$. Negative strain corresponds to in-plane compressive strain, while positive strain corresponds to in-plane tensile strain. Blue indicates the orbital contribution from Ni $3d_{x^2-y^2}$, and red represents the contribution from Ni $3d_{z^2}$. The inset in the last panel of \textbf{A} illustrates the Brillouin zone and the high-symmetry momentum path used in the calculations. Electronic structure calculations are implemented with DFT+$U$ with $U=3.5$eV.}
\label{fig:band}
\end{figure*}

\subsection*{Electronic Structure Properties}

The electronic structure of strained bilayer nickelates Re$_3$Ni$_2$O$_7$ for all lanthanide element Re are calculated using first-principles DFT+$U$ calculations with $U=3.5$eV. For unstrained La$_3$Ni$_2$O$_7$, the prominent features near Fermi energy are the bonding ($\gamma$) and anti-bonding ($\gamma_{an}$) bands at $\Gamma$ point with Ni $d_{z^2}$ orbital character, and $\beta$ bands at M point with primarily Ni $d_{x^2-y^2}$ orbital character, as shown in Fig. \ref{fig:band}\textbf{A}. Such features are consistent with other first-principles studies \cite{UValueyang2024orbital,Geisler2024,PhysRevLett.132.106002,Wang2024} for La$_3$Ni$_2$O$_7$, and persistent for Pr$_3$Ni$_2$O$_7$ and Nd$_3$Ni$_2$O$_7$, as shown in Fig. \ref{fig:band}\textbf{B} and \textbf{C}.

Under in-plane biaxial compressive strain, overall, the bands associated with the Ni $3d_{z^2}$ orbitals shift downward relative to the Fermi level and become flatter, while those associated with the Ni $3d_{x^2-y^2}$ orbitals shift upward and become more dispersive. This trend is reflected in the downshift of the $\gamma$ and $\gamma_{an}$ bands and the upshift of the $\beta$ bands under 2.5\% compressive strain. These shifts become more pronounced at 5\% compressive strain. On the other hand, under tensile strain, the trend is reversed: the Ni $3d_{z^2}$ bands shift upward and become more dispersive, while the Ni $3d_{x^2-y^2}$ bands shift downward and become narrower. Specifically, the $\gamma$ band shifts to higher energy and goes across the Fermi energy, and the bottom of the $\beta$ bands shift lower with increased $3d_{x^2-y^2}$ weight, at 2.5\% tensile strain. Such trends are persistent for La$_3$Ni$_2$O$_7$, Pr$_3$Ni$_2$O$_7$, and Nd$_3$Ni$_2$O$_7$, as shown in Fig. \ref{fig:band}.

%Previous studies have shown that in La$_3$Ni$_2$O$_{7-\delta}$, oxygen deficiencies ($\delta$) exceeding 0.08 lead to an insulating state~\cite{doi:10.1143/JPSJ.64.1644}. Likewise, in thin films, the absence of an ozone annealing procedure results in insulating behavior~\cite{Ko2024,liu2025superconductivitynormalstatetransportcompressively}. Research suggests that oxygen vacancies predominantly occur at the apical inner sites, significantly suppressing superconductivity~\cite{zhang2023work2,bhatt2025resolvingstructuraloriginssuperconductivity}. These observations highlight the critical role of hybridization between Ni \(3d\) orbitals and O \(2p\) orbitals in governing both the normal-state transport properties and the underlying mechanism of superconductivity.

%Since transport and superconductivity primarily involve the bands around the Fermi level, which predominantly consist of Ni \(3d_{x^2-y^2}\) and \(3d_{z^2}\) orbitals, theoretical studies have identified the Ni \(3d_{z^2}\) orbital as the main contributor to superconducting \(s\pm\) pairing~\cite{Luo2024,PhysRevLett.131.126001,PhysRevLett.132.106002,Lu2024,PhysRevB.109.045154}. This raises the intriguing question of the role played by the oxygen \(2p\) orbital in these bands and how its contribution evolves under strain. To address this, 
We analyzed the orbital components of the $\gamma$ and $\gamma_{an}$ bands at the $\Gamma$ point, which are primarily composed of Ni-\(3d_{z^2}\) and O-\(2p_z\) orbitals. For the $\gamma$ bands, the O-\(2p_z\) orbital contribution predominantly originates from the apical outer oxygen atoms. Conversely, for the $\gamma_{an}$ bands, the O-\(2p_z\) orbital contribution primarily comes from the apical inner oxygen atoms. Under 2.5\% compressive strain, the O-$2p_z$ orbital contributions in the two $\gamma$ bands and the two $\gamma_{an}$ bands closest to the Fermi energy increase by approximately 10\% compared to the unstrained case. Detailed hybridization data for all eight relevant bands are provided in the Supplementary Materials.

\subsection*{Tight-binding Parameters of the Strained Structure in Amam Phase}

\begin{figure*}[htp!]
  \centering  
\includegraphics[width=\columnwidth]{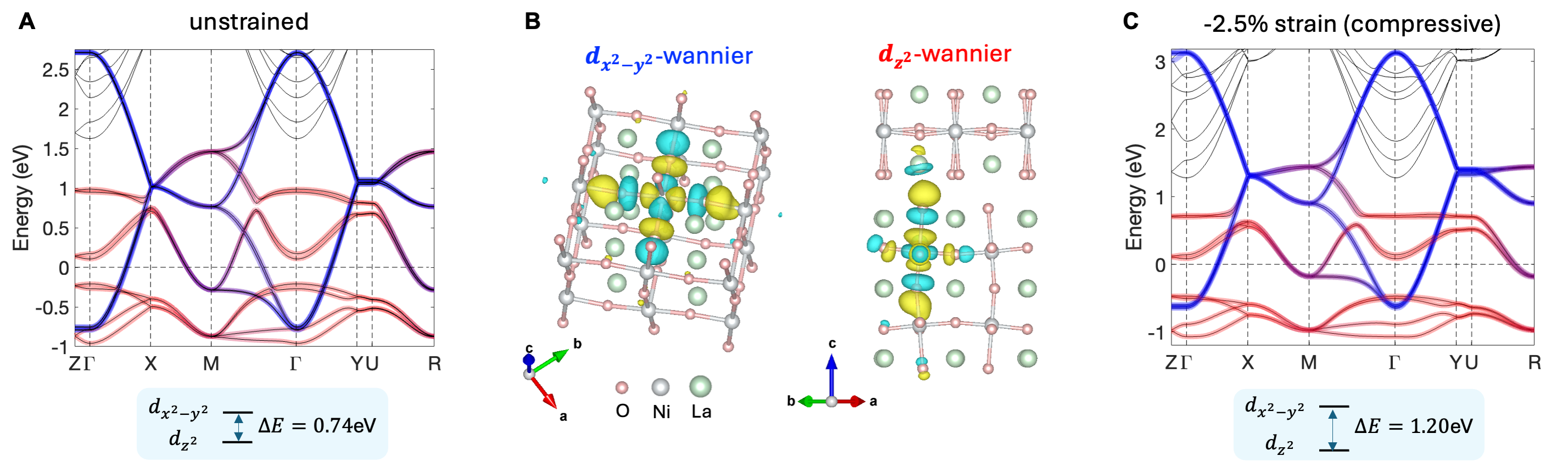}
\caption{\textbf{Wannier downfolding of electronic structure for unstrained and strained bilayer nickelate La$_3$Ni$_2$O$_7$. A}, DFT+$U$ band structure (black) compared with the band structure obtained from a sixteen-orbital tight-binding model via Wannier downfolding using Ni $3d_{x^2-y^2}$ and $3d_{z^2}$ Wannier orbitals (blue and red) for unstrained La$_3$Ni$_2$O$_7$. Contributions from the Ni $3d_{x^2-y^2}$
  Wannier orbitals are shown in blue, while contributions from Ni $3d_{z^2}$ Wannier orbitals are shown in red. \textbf{B}, Visualization of the Ni $3d_{x^2-y^2}$ and Ni $3d_{z^2}$ Wannier orbitals for unstrained La$_3$Ni$_2$O$_7$. The Wannier orbitals for -2.5\% strained material are almost visually indistinguishable from the unstrained case. \textbf{C}, DFT+$U$ band structure (black) compared with the band structure from the sixteen-orbital tight-binding model derived via Wannier downfolding (blue and red) for La$_3$Ni$_2$O$_7$ under 2.5\% compressive strain.  }
  \label{fig:Fig3}
\end{figure*}

To investigate the microscopic origin of the observed changes in the band structure under strain, we constructed a low symmetry($Amam$) tight-binding model and obtained the precise tight-binding parameters using post-DFT Wannier downfolding calculations for both the unstrained and 2.5\% compressive strain cases of La$_3$Ni$_2$O$_7$. The Wannier downfolding was performed using a sixteen-orbital model, considering the Ni $d_{x^2-y^2}$ and $d_{z^2}$ Wannier orbitals from all eight Ni atoms in the unit cell of the orthorhombic \textit{Amam} phase. This procedure was implemented using Wannier90~\cite{Pizzi2020}, interfaced with Quantum Espresso~\cite{QE-2017} for the band structure calculations. The Ni 3$d_{x^2-y^2}$ bands are entangled with bands predominantly composed of La orbitals due to energy overlap above 1 eV. To isolate the target bands associated with the Ni 3$d_{x^2-y^2}$ and 3$d_{z^2}$ orbitals, we employed the disentanglement procedure~\cite{PhysRevB.65.035109} implemented in Wannier90, which ensures a more accurate representation of the low-energy tight-binding description. The resulting band structure from the sixteen-orbital tight-binding model, obtained via Wannier downfolding, matches perfectly with the DFT bands, as shown in Fig.~\ref{fig:Fig3}\textbf{A} and \textbf{C} for the unstrained and 2.5\% compressive strain cases. The obtained Wannier orbitals for one Ni atom are visualized in Fig.~\ref{fig:Fig3}\textbf{B}. The key tight-binding model parameters are listed in Table~\ref{hoppingwan}, and the tight-binding Hamiltonian is provided in the Methods section.

 \begin{table}[h!]
 \large
 \captionsetup{justification=raggedright, singlelinecheck=false, width=\columnwidth}
\centering
 \begin{tabular}{|c|c|c|c|c|}
  \hline\hline
 % \multicolumn{3}{|c|}{Hopping parameters} \\
 % \hline
  & Unstrained  & 2.5\% Compress & 29.5 GPA~\cite{PhysRevLett.131.126001}& 50 GPA\cite{Zhang2024}\\
   \hline\hline
    $\epsilon^{x}$& 0.848   & 1.034 & 0.776& 0.551\\
   \hline
    $\epsilon^{z}$& 0.106   & -0.165 & 0.409& 0\\
   \hline
  $t_{1}^x$($t_{1'}^{x}$)& -0.419   & -0.460 (-0.443) & -0.483&-0.559\\
   \hline
   $t_{1}^z$ ($t_{1'}^z$)&-0.0996 (-0.0675)  & -0.0828 (-0.0445) & -0.110&-0.125\\
   \hline
   $t_{2}^x$ ($t_{2'}^x$)& 0.0625 (0.0808) & 0.0667 (0.0868) & 0.069&NA\\
   \hline

   $t_{2}^z$& -0.0138  &-0.0116 & -0.017&NA\\
   \hline
   $t_{3}^z$& -0.0447  &-0.0495 &NA &NA\\
   \hline
   $t_{1}^{xz}$($t_{1'}^{xz}$)&-0.208 (-0.181)  & -0.205 (-0.165) & -0.239&-0.270\\
   \hline
      
   $t_{\bot}^x$& 0.008 & 0.007 & 0.005&NA\\
   \hline  
   $t_{\bot}^z$& -0.594 & -0.631 & -0.635&-0.719\\
   \hline  
   $t_{1\bot}^{xz}$& 0.0265 & 0.03 & 0.034&NA\\
   \hline\hline 
 \end{tabular}

\caption{\textbf{Tight-binding parameters for bilayer nickelates under different conditions: unstrained, with 2.5\% biaxial in-plane compressive strain, at 29.5 GPa, and at 50 GPa (all values in eV).}  Here, $\epsilon^x$ and $\epsilon^z$ denote the on-site energies for the $d_{x^2-y^2}$ and $d_{z^2}$ orbitals, respectively. The hopping parameters are indexed as follows: the superscript $x$ indicates hopping between two $d_{x^2-y^2}$ orbitals, $z$ denotes hopping between two $d_{z^2}$ orbitals, and $xz$ represents hopping between $d_{x^2-y^2}$ and $d_{z^2}$ orbitals. The subscript $\bot$ refers to inter-layer hopping, while subscripts 1, 2, and higher denote intra-layer hopping between nearest neighbors, next-nearest neighbors, etc. If hopping along the two in-plane directions is non-degenerate, primed indices (e.g., $1'$, $2'$) are used to differentiate them. The tight-binding parameters for the 29.5 GPa and 50 GPa cases are obtained from Wannier downfolding of DFT calculations, as reported in Refs.~\cite{PhysRevLett.131.126001} and \cite{Zhang2024}, respectively.}

\label{hoppingwan} 
 \end{table}

For the unstrained case, the crystal field splitting of the on-site energy between the $d_{x^2-y^2}$ and $d_{z^2}$ orbitals is $\epsilon^{x} - \epsilon^{z} = 0.74$~eV, as shown in Fig.~\ref{fig:Fig3}\textbf{A} and Table 1. Under 2.5\% compressive strain, this on-site energy difference $\epsilon^{x}-\epsilon^{z}$ increases to 1.20~eV, which is consistent with the downward shift of the $d_{z^2}$ bands and the upward shift of the $d_{x^2-y^2}$ bands discussed in the previous section. %Such a change may lead to a greater occupation of the $d_{z^2}$ orbital compared to the $d_{x^2-y^2}$ orbital under compressive strain. 
The on-site energy difference $\epsilon^{x} - \epsilon^{z}$ for both unstrained and 2.5\% compressive strained cases are significantly larger than the 0.37~eV reported for the Wannier downfolding in the DFT calculation of the {\it Fmmm} phase at 29.5 GPA~\cite{PhysRevLett.131.126001}, and it also exceeds the 0.551~eV obtained from the Wannier downfolding of the DFT+$U$ calculation for the {\it Fmmm} phase at 50 GPA~\cite{Zhang2024}.  %Since the electrons in the $d_{z^2}$ orbital are recognized as the dominant contributors to $s\pm$ pairing~\cite{Luo2024}, this change may enhance superconductivity.

Under 2.5\% compressive strain, the in-plane hopping between nearest-neighbor $d_{z^2}$ orbitals, denoted as $t_{1}^z$ (or $t_{1'}^z$ along the other basal direction), is reduced by one-third along one direction (about 17\% along the other), compared to the unstrained case. This reduction decreases the bandwidth of the $d_{z^2}$ bands, affecting both the inter-layer bonding band below the Fermi level ($\gamma$ as shown in Fig. \ref{fig:band}\textbf{A}) and inter-layer anti-bonding band above the Fermi level ($\gamma_{an}$ as shown in Fig. \ref{fig:band}\textbf{A}). Interestingly, under 2.5\% compressive strain, the magnitude of $t_{1}^z$ is smaller along both in-plane Ni-Ni directions compared to the high-symmetry {\it Fmmm} phase at a high pressure of 29.5 GPa, which is counterintuitive.  %As a result, the $d_{z^2}$ bands become flatter, as shown in Figure~\ref{fig:Fig3}. 

Another difference between the strained and unstrained cases is the change in \( t_{\bot}^z \), which is the inter-layer hopping between the two Ni-$3 d_{z^2}$ orbitals via the O-$2p_z$ orbitals of the intermediate apical oxygen. Despite the vertical elongation of the lattice under compressive strain, the increase in \( t_{\bot}^z \) may result from both the enhanced apical Ni-O-Ni bond angle and larger Wannier orbital spread. This increase in \( t_{\bot}^z \) leads to a larger energy separation between the inter-layer bonding states ($\gamma$ in Fig. \ref{fig:band}\textbf{A} below the Fermi level) and the anti-bonding states ($\gamma_{an}$ in Fig. \ref{fig:band}\textbf{A} above the Fermi level). Under 2.5\% compressive strain, \( t_{\bot}^z \) approaches the reported value of 0.635 for the high-symmetry {\it Fmmm} phase under 29.5GPA~\cite{PhysRevLett.131.126001}.

%\textcolor{blue}{CJ: the discussion about t3x may be removed.}By analyzing the maximally-localized Wannier functions (MLWFs) from the Wannier90 calculations, we found that the spreads of the MLWFs are on the order of several angstroms. Notably, the spread of the $d_{x^2-y^2}$ MLWF is larger than that of $d_{z^2}$, which results in a significantly large $t_{3x}$. This parameter, $t_{3x}$, contributes to a 0.2 eV band splitting at the X-point between the $d_{x^2-y^2}$ and $d_{z^2}$ bands.

For the hopping parameters of the Ni $d_{x^2-y^2}$ orbital, a 2.5\% in-plane compressive strain increases all intra-layer hopping amplitudes, including nearest-neighbor, next-nearest-neighbor, and next-next-nearest-neighbor hoppings, by 6\% to 10\% relative to the unstrained case. This enhancement corresponds to the observed increase in band dispersion for the 
$d_{x^2-y^2}$ bands and aligns with the expectation that reducing in-plane lattice parameters strengthens hopping amplitudes for the Ni $d_{x^2-y^2}$ orbital. Compared to the in-plane hopping parameters for Ni $d_{x^2-y^2}$ in the high-symmetry {\it Fmmm} phase under high pressure (29.5 GPa), the nearest-neighbor hoppings for Ni $d_{x^2-y^2}$ for 2.5\% compressive strain are smaller by 5\% and 8\% along the two in-plane Ni-Ni directions, respectively.

We note that the Wannier downfolding was performed on the band structure derived from DFT+$U$ calculations with \( U = 3.5 \, \text{eV} \), in order to reproduce the experimental ARPES-measured band structure for the unstrained material. Consequently, the tight-binding parameters obtained here may differ from those derived from DFT calculations without the \( U \) correction, though the difference is expected to be minimal. Furthermore, additional many-body calculations that account for strong correlation effects, based on the tight-binding model parameters we propose, need to be carefully implemented to avoid double-counting of correlation effects.

\setcounter{subfigure}{0}

\section*{Discussions}

Bilayer nickelates have exhibited superconductivity under both high pressure~\cite{80KSuper} and epitaxial compressive strain~\cite{Ko2024}, with each method inducing distinct modifications to the material’s structural and electronic properties. By comparing our first-principles calculations for the compressive strain case in the {\it Amam} phase with the structural and electronic characteristics of the high-pressure phase, as investigated in previous studies~\cite{Geisler2024,PhysRevLett.131.126001,Zhang2024,PhysRevLett.131.236002}, we gain deeper insights into the key material properties that may play a crucial role in the emergence of superconductivity.

Our calculations suggest that Ni \( d_{z^2} \) bands may not need to cross the Fermi energy for superconductivity to emerge. In the high-pressure {\it Fmmm} phase, theoretical calculations have shown that the bonding \( 3d_{z^2} \) band shifts above the Fermi surface at the \(\Gamma\) point, leading to the formation of a new hole pocket~\cite{Geisler2024,PhysRevLett.131.126001}. However, our calculations for the compressive strain case reveal that the bonding \( 3d_{z^2} \) bands shift further downward, away from the Fermi energy, due to a significant decrease in the \( 3d_{z^2} \) site energy, thereby preventing the formation of a new hole pocket. The recent experimental realization of superconductivity in epitaxially compressively strained bilayer nickelate La$_3$Ni$_2$O$_7$~\cite{Ko2024}, combined with the absence of $3d_{z^2}$ bands near the Fermi level in our calculations, suggests that 
$3d_{z^2}$ bands may not play a critical role in the superconductivity of bilayer nickelates.

% Apical Ni-O-Ni bond angle may not need to reach precisely 180$^\circ$ to achieve the emergence of superconductivity, due to our DFT+$U$ calculations for Ni-O-Ni bond angle increases to a maximum of 171.1\(^\circ\) upon compressive strain in the \textit{Amam} phase, and the realization of superconductivity in epitaxial compressively strained bilayer nickelate La$_3$Ni$_2$O$_7$~\cite{Ko2024}. A recent experimental research demonstrted that the superconductivity under high pressure is achieved in orthorhombic structure\cite{shi2025prerequisitesuperconductivitysdwtetragonal}. Another indication is that superconductivity transition can be achieved within the orthorhombic {\it Amam} phase around 7 to 11 GPA for La$_2$PrNi$_2$O$_7$, where the material is in orthorhombic {\it Amam} phase where the apical Ni-O-Ni angle has not reached 180$^\circ$ yet~\cite{wang2024nature}. Additionally, La$_2$PrNi$_2$O$_7$ thin films show increased {\it T}$_c$ onset and {\it T}$_c$ zero\cite{liu2025superconductivitynormalstatetransportcompressively} compared with La$_3$Ni$_2$O$_7$ thin films\cite{Ko2024}, and the substitution of La with Pr in bilayer nickelates will decrease the apical Ni-O-Ni angle away from 180$^\circ$ according to our calculations.

Our calculations show that the apical Ni--O--Ni bond angle may not need to reach exactly 180$^\circ$ to facilitate the emergence of superconductivity. Our DFT+$U$ calculations indicate that under 2.5\% compressive strain within the \textit{Amam} phase — closely matching experimental conditions for strain-induced superconductivity~\cite{Ko2024} — the Ni--O--Ni bond angle increases to 171.1$^\circ$ but does not reach 180$^\circ$. Also, in La$_2$PrNi$_2$O$_7$, the superconducting transition takes place within the orthorhombic \textit{Amam} phase at pressures between 7 and 11~GPa, where the apical Ni--O--Ni bond angle has not yet reached 180$^\circ$~\cite{wang2024nature}. Additionally, thin films of La$_2$PrNi$_2$O$_7$ exhibit higher $T_{c,\text{onset}}$ and $T_{c,\text{zero}}$ compared to La$_3$Ni$_2$O$_7$ thin films~\cite{liu2025superconductivitynormalstatetransportcompressively, Ko2024}. But, according to our calculations, substituting La with Pr in bilayer nickelates further reduces the apical Ni--O--Ni bond angle, suggesting that this apical angle may not be critical for superconductivity.

The on-site energy difference between the $3d_{x^2-y^2}$ and $3d_{z^2}$ orbitals is the primary parameter modulated by strain and may play a crucial role in tuning superconductivity in bilayer nickelates — an aspect that has been largely overlooked in previous studies. In the high pressure {\it Fmmm} phase, the site energy (crystal field) difference between the $d_{z^2}$ and $d_{x^2-y^2}$ orbitals is relatively small. However, under strain, this difference increases significantly, with the site energy of the $d_{z^2}$ orbital decreasing. Consequently, the $d_{z^2}$ bands shift downward, and their density of states (DOS) peak moves below the Fermi level, leading to an increased occupation of the $d_{z^2}$ orbital, as shown in Fig.~S3 of the Supplementary Material.

% Theoretical studies \cite{Lu2024, Luo2024,PhysRevLett.131.236002} suggest that single-orbital bilayer $t-J-J_\bot$ model is the low-energy effective model for bilayer nickelates with superconductivity at high-pressure with relatively large $J_\bot/J$ favors a $s$-wave pairing superconductivity. Other suggested a two-orbital bilayer $t-J-J_\bot$ model and shows that without doping, the $s\pm$ pairing mechanism dominates the superconductivity, with the primary contribution coming from $d_{z^2}$ electrons\cite{Luo2024,PhysRevLett.131.126001,PhysRevB.109.045154,Zhang2024,PhysRevB.110.094509}. Both inter-layer and intra-layer exchange interactions play significant roles in determining the pairing order parameter. Therefore, an increased electron population in the $d_{z^2}$ orbital could enhance the $s\pm$ pairing and favor superconductivity. Taking $J\approx 4t^2/U$, our studies show that for unstrained case and 2.5\% compressive strained case, $J_\bot/J$ is 2.00 and 1.95, while for 29.5GPA $J_\bot/J$ is 1.73. Taking the on-site energy difference as an potentially important factor, two-orbital bilayer $t-J-J_\bot$ model or two-orbital bilayer Hubbard model, which accounts for the on-site energy difference and electron occupation disparity, will be more likely the microscopic model for bilayer nickelates.

Theoretical studies~\cite{Lu2024, Luo2024, PhysRevLett.131.236002} suggest that the single-orbital bilayer $t$-$J$-$J_\perp$ model may serve as the low-energy effective model for bilayer nickelates with superconductivity at high pressure. In this framework, a relatively large $J_\perp/J$ favors $s$-wave pairing superconductivity. Other studies have proposed a two-orbital bilayer $t$-$J$-$J_\perp$ model, demonstrating that in the undoped regime, the $s\pm$ pairing mechanism dominates superconductivity, with the primary contribution originating from $d_{z^2}$ electrons~\cite{Luo2024, PhysRevLett.131.126001, PhysRevB.109.045154, Zhang2024, PhysRevB.110.094509}. Both interlayer and intralayer exchange interactions play crucial roles in determining the pairing order parameter. Therefore, an increased electron population in the $d_{z^2}$ orbital could enhance $s\pm$ pairing and promote superconductivity.  
Using $J \approx 4t^2/U$, our calculations yield $J_\perp/J$ values of 2.00 and 1.95 for the unstrained and 2.5\% compressively strained cases, respectively, while for the 29.5~GPa high-pressure case, $J_\perp/J$ is 1.73. Given the significance of the on-site energy difference as a potentially critical factor, a two-orbital bilayer $t$-$J$-$J_\perp$ model or a two-orbital bilayer Hubbard model—accounting for both the on-site energy difference and electron occupation disparity—would likely provide a more accurate microscopic description of bilayer nickelates.

We note that the tight-binding model parameters for the unstrained case do not differ significantly from those of the compressive strained and high-pressure cases, except for the on-site energy difference. This raises an important question: Is the emergence of superconductivity primarily driven by intrinsic modifications to the material's properties induced by strain and pressure, or is it largely influenced by extrinsic factors, such as the purity of the bilayer phase or the presence of oxygen vacancies? Indeed, recent experimental studies suggest that oxygen concentration may play a more critical role than structural transitions in enabling superconductivity in bilayer nickelates~\cite{shi2025prerequisitesuperconductivitysdwtetragonal}. Further studies focusing on the synthesis of high-quality bilayer nickelate samples may help clarify this issue.

\section*{Methods}
\subsection*{First-principles DFT calculations}
To investigate the structural and electronic properties of non-magnetic $\text{Re}_3$$\text{Ni}_2$$\text{O}_7$ under strain, we performed first-principles density functional theory (DFT) calculations. The lattice parameters and atomic positions of the unstrained structures were fully relaxed, using the Vienna {\it ab-initio} Simulation Package (VASP) \cite{PhysRevB.47.558}. For the strained cases,  the relaxed lattice parameters, atomic positions and the corresponding band structures were obtained using VASP as well.

We used projector augmented-wave (PAW) pseudopotentials, treating the 
$f$-electrons of the rare-earth elements as core states, and employed the generalized gradient approximation (GGA) with the Perdew-Burke-Ernzerhof (PBE) exchange-correlation functional \cite{pbeexchange}. To account for the electronic correlations, the Dudarev formulation of the on-site Hubbard $U$ correction was applied to all Ni sites\cite{dudarevU}, with a $U$ value of 3.5 eV, consistent with previous studies \cite{geislerUvalue, UValueyang2024orbital}.

The convergence criterion for ionic relaxation (EDIFFG) was set to \(8 \times 10^{-5}\,\mathrm{eV}\) for La\(_3\)Ni\(_2\)O\(_7\), and \(2 \times 10^{-2}\,\mathrm{eV}\) for other Re\(_3\)Ni\(_2\)O\(_7\) compounds. The electronic energy convergence threshold (EDIFF) was specified as \(8 \times 10^{-6}\,\mathrm{eV}\). The Methfessel-Paxton smearing method, with a smearing width of 0.2\,eV, was employed to account for metallic electronic states. A plane-wave kinetic energy cutoff of 520\,eV was used for the basis set. The Brillouin zone was sampled using a Monkhorst-Pack grid of \(8 \times 8 \times 2\)~\cite{Monkhorst}.

\subsection*{Tight-binding Hamiltonian}
The tight-binding model considered here includes 4 Ni atoms or 8 orbitals, with the basis defined as 

\[
\Psi = \big(d_{A\alpha x}, d_{A\alpha z}, d_{A\beta x}, d_{A\beta z}, d_{B\alpha x}, d_{B\alpha z}, d_{B\beta x}, d_{B\beta z}\big)^T
\]
where \(A\) and \(B\) denote the bottom and top layers, respectively, \(\alpha\) and \(\beta\) refer to the two Ni atoms within the same layer, \(x\) corresponds to the \(3d_{x^2-y^2}\) orbital, and \(z\) corresponds to the \(3d_{z^2}\) orbital.
This basis effectively captures the essential degrees of freedom required to model the electronic structure of the bilayer system, accounting for both intra-layer and inter-layer couplings. Although our Wannier downfolding calculation includes 16 bands, as shown in Figure~\ref{fig:Fig3}, this is due to the unit cell being a supercell containing two bilayer structures. However, the coupling between these two bilayers is minor, resulting in almost degenerate bands that reduces the 16 bands to 8 distinct bands.

% as well as layer-dependent hybridization effects. 

The Hamiltonian of the eight-orbital tight-binding model is written as:

\begin{multline}
\hfill       H(k)=
  \begin{bmatrix} 
  H_A(k) & H_{AB}(k) 
  \\ H^{\ast}_{AB}(k)  & H_B(k)
  \end{bmatrix} \hfill\\
  \\
   H_A(k)=H_B(k)=
  \begin{bmatrix}
T^x(k)      &  0            & V^x(k)        &V^{xz}(k) \\
0           &T^z(k)         &V^{xz}(k)    & V^z(k)\\
V^{x*}(k)   &V^{xz*}(k)     &T^x(k)         & 0\\
V^{xz*}(k)  &V^{z*}(k)     &0              &T^z(k)
  \end{bmatrix}\\
  \\
 \hfill  H_{AB}(k)=
    \begin{bmatrix}
    t_{\bot}^x  &0              &0      &V_{1\bot}^{xz}(k)\\
    0           &t_{\bot}^z     &V_{1\bot}^{xz}(k)      &0\\
    0           &V_{1\bot}^{xz}(k)&t_{\bot}^x           &0&\\
V_{1\bot}^{xz}(k)   &0          &0      &  t_{\bot}^z\\
    \end{bmatrix}\hfill
\end{multline}

\begin{equation}
\begin{aligned}
    T^x(k) &=\epsilon^x
+2t_{2}^x\text{cos}(k_xa)+2t_{2'}^x\text{cos}(k_y b)
    +4t_{3}^x\text{cos}(k_xa)\text{cos}(k_y b)\\
   T^z(k) &=\epsilon^z+2t_{2}^z(\text{cos}(k_xa)+\text{cos}(k_y b))
\end{aligned}
\end{equation}
% \begin{equation}
% \begin{split}
%     \epsilon_{\alpha x}'=\epsilon_{\alpha x}
% +2t_{2x}cos(k_xa)+2t'_{2x}cos(k_y b)\\
%     +4t_{3x}cos(k_xa)cos(k_y b)
% \end{split}   
% \end{equation}
% \begin{equation}
%     \epsilon_{\alpha z}'=\epsilon_{\alpha z}+2t_{2z}cos(k_xa)+2t_{2z}cos(k_y b)
% \end{equation}
Under ambient pressure, the lattice parameters are $a=5.367$ \AA, $b=5.44$ \AA.
% \begin{equation}
%     k_y=k_c
% \end{equation}
% \begin{equation}
%     k_x=\Vec{k}_a-\Vec{k}_b
% \end{equation}

\begin{equation}
\begin{aligned}
    V^x(k) &=t_{1}^x(e^{ik\cdot\Vec{R}_{\alpha\beta}}+e^{ik\cdot\Vec{R}_{\alpha\beta2}})+t_{1'}^x(e^{ik\cdot\Vec{R}_{\alpha\beta1}}+e^{ik\cdot\Vec{R}_{\alpha\beta3}})
\\V^z(k) &=t_{1}^z(e^{ik\cdot\Vec{R}_{\alpha\beta}}+e^{ik\cdot\Vec{R}_{\alpha\beta2}})+t_{1'}^z(e^{ik\cdot\Vec{R}_{\alpha\beta1}}+e^{ik\cdot\Vec{R}_{\alpha\beta3}})\\
     V^{xz}(k) &=t_{1}^{xz}(e^{ik\cdot\Vec{R}_{\alpha\beta}}-e^{ik\cdot\Vec{R}_{\alpha\beta2}})+t_{1'}^{xz}(e^{ik\cdot\Vec{R}_{\alpha\beta1}}-e^{ik\cdot\Vec{R}_{\alpha\beta3}})\\
   V_{1\bot}^{xz}(k) &=t_{1\bot}^{xz}(e^{ik\cdot\Vec{R}_{\alpha\beta}}+e^{ik\cdot\Vec{R}_{\alpha\beta1}}-e^{ik\cdot\Vec{R}_{\alpha\beta2}}-e^{ik\cdot\Vec{R}_{\alpha\beta3}})
    \end{aligned}\label{hopping}
\end{equation}

In this context, $\Vec{R}_{\alpha\beta}$ and $\Vec{R}_{\alpha\beta2}$ are vectors representing the distances between the two nearest Ni atoms along the basal 1 direction, while $\Vec{R}_{\alpha\beta1}$ and $\Vec{R}_{\alpha\beta3}$ correspond to those along the basal 2 direction. For the hopping parameter index, the number 1 indicates nearest-neighbor (NN) hopping, 2 represents next-nearest-neighbor hopping, and so on. The symbol $x$ denotes hopping between two $d_{x^2-y^2}$ orbitals, $z$ denotes hopping between two $d_{z^2}$ orbitals, and $xz$ denotes hopping between $d_{x^2-y^2}$ and $d_{z^2}$ orbitals. The symbol $\bot$ indicates inter-layer hopping.  

Due to the reduced symmetry of the {\it Amam} phase, the hopping parameters vary between different directions. In Eq.~\ref{hopping}, the values with subscript 1 represent nearest-neighbor hopping along the Ni-O basal 2 direction (shorter bond length), while the values with subscript $1'$ correspond to hopping along the basal 1 direction, as illustrated in Fig.~\ref{bar}. Additionally, subscript 2 denotes next-nearest neighbor hopping along the lattice vector \(a\), and subscript $2'$ corresponds to next-nearest neighbor hopping along the lattice vector \(b\).

\section*{Acknowledgments}
We thank Benjamin Geisler and Bai Yang Wang for insightful discussions. This work is supported by the Center for Molecular Magnetic Quantum Materials, an Energy Frontier Research Center funded by the U.S. Department of Energy,
Office of Science, Basic Energy Sciences under Award no.
DE-SC0019330. The work at Stanford is supported by the U.S. Department of Energy, Office of Science,
Basic Energy Sciences, Division of Materials Sciences and Engineering,
under contract No. DE-AC02-76SF00515. Computations were done using the utilities of the University of Florida Research Computing.
\bibliography{main}

\vspace{12cm}
\begin{center}
   \LARGE Supplementary Materials 
\end{center}
\maketitle
\setcounter{figure}{0}
\setcounter{table}{0}
\renewcommand{\theequation}{S\arabic{equation}}
\renewcommand{\thefigure}{S\arabic{figure}}
\section{More details on DFT calculations}
We began with the experimentally determined crystal structure of $\text{La}_3\text{Ni}_2\text{O}_7$ in the orthorhombic {\it Amam} phase under ambient conditions~\cite{ExpCrystalStruc}. Lanthanum (La) was systematically substituted with each element from the lanthanide series, and the resulting unit cells were fully relaxed to obtain their unstrained equilibrium structures. For each rare-earth element (Re), we then optimized the structure under four distinct in-plane strain conditions: 2.5\% compressive strain ($-2.5\%$), 5\% compressive strain ($-5\%$), 2.5\% tensile strain ($+2.5\%$), and 5\% tensile strain ($+5\%$). 

For the compressive strain conditions ($-2.5\%$ and $-5\%$), the in-plane lattice parameters (\textbf{a} and \textbf{b}) were reduced by 2.5\% and 5\%, respectively. Conversely, for the tensile strain conditions ($+2.5\%$ and $+5\%$), \textbf{a} and \textbf{b} were increased by the corresponding percentages. In all cases, the out-of-plane lattice parameter (\textbf{c}) was scanned in 0.1~Å increments, with atomic positions fully relaxed at each step. The final strained configuration for each case was determined as the structure with the lowest total energy. Using this approach, the optimized crystal structures under all strain conditions were systematically obtained.

\section{Basal Ni-O-Ni bond angles and Ni-O bond lengths}
\begin{figure}[htbp]
\centering
\includegraphics[width=0.9\linewidth]{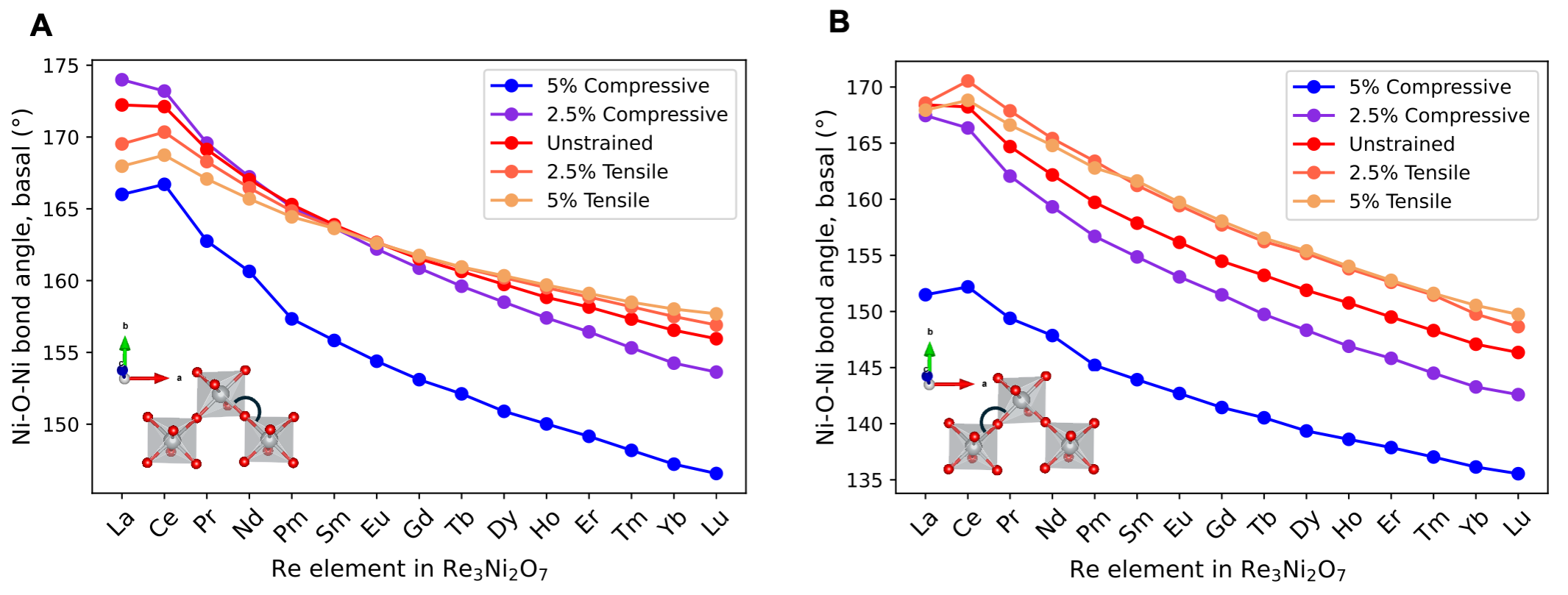}
\caption{\large\textbf{Variation of in-plane Ni-O-Ni bond angles under different strain conditions across the lanthanide series.} 
\textbf{A} Evolution of the basal Ni-O-Ni bond angle along one in-plane direction with applied strain. The inset provides a schematic illustration of the corresponding basal octahedral angle. 
\textbf{B} Variation of the basal Ni-O-Ni bond angle along the orthogonal in-plane direction to that shown in \textbf{A}. The inset depicts a visualization of the basal octahedral angle for this direction.}
\label{s1}
\end{figure}

The in-plane Ni-O-Ni bond angles exhibit a distinct trend under strain compared to high-pressure conditions. Under high pressure, the bond angles along two orthogonal directions increase uniformly toward 180$^\circ$. In contrast, strain induces an anisotropic response: one bond angle decreases, while the other increases, as illustrated in Figure~\ref{s1}. This divergence underscores the anisotropic structural deformation caused by in-plane strain, in stark contrast to the isotropic behavior observed under high pressure. 

\begin{figure}[htbp]
\centering
\includegraphics[width=0.9\linewidth]{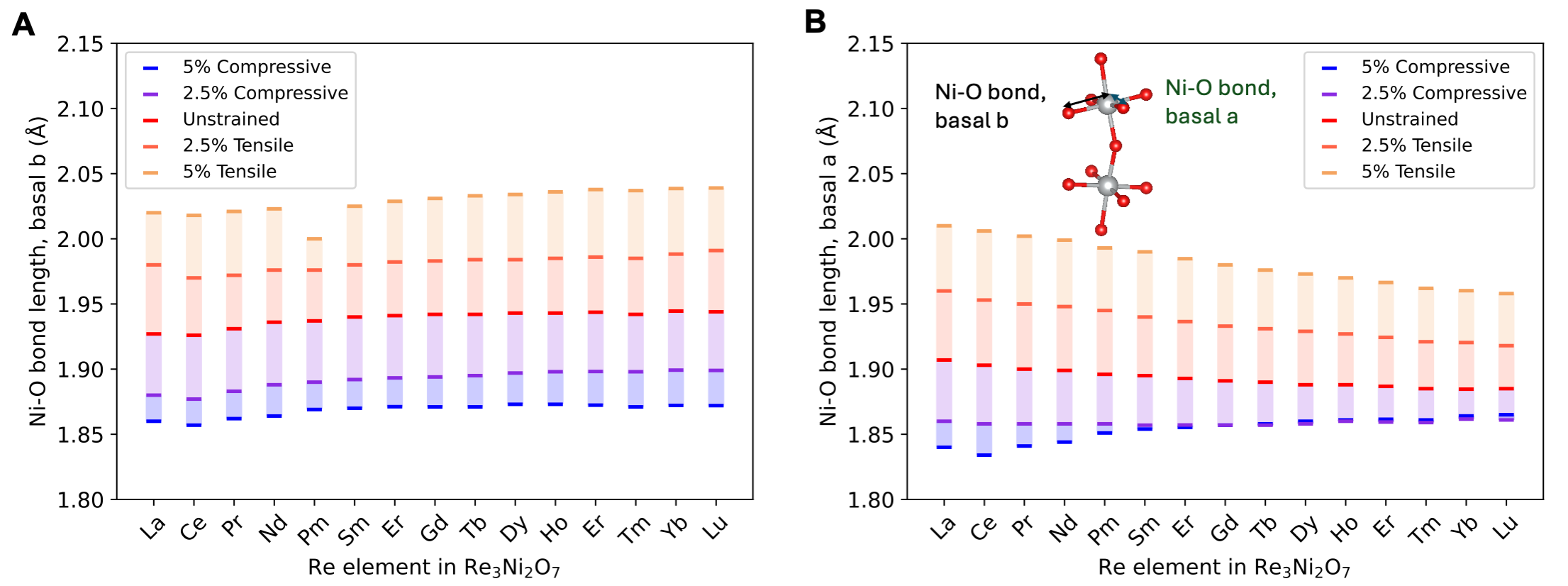}

\caption{\large\textbf{Variation of in-plane Ni-O bond lengths under different strain conditions across the lanthanide series.} 
\textbf{A} Evolution of the in-plane Ni-O bond length along the basal \(b\) with applied strain. 
\textbf{B} Variation of the in-plane Ni-O bond length along the basal \(a\) with applied strain. The inset provides a schematic illustration of the basal \(a\) and \(b\) directions.}

% \caption{\textbf{Variation of in-plane Ni-O bond length under different strain conditions across the lanthanide series.}\textbf{A}Evolution of the basal Ni-O bond legnth along one in-plane direction basal b with applied strain.\textbf{B} Variation of the basal Ni-O bond length along the in-plane direction basal a. The inset depicts a visualization of the basal a and b direction.}
\label{s2}
\end{figure}

A similar anisotropic trend is observed in the in-plane Ni-O bond lengths along the basal \(a\) and \(b\) axes, as Figure~\ref{s2} shows. Under compressive strain, both bond lengths decrease, but the bond length along the basal \(b\) axis becomes slightly shorter than that along the basal \(a\) axis. In comparison, under high pressure, the bond lengths along the two directions remain equal, further emphasizing the different structural responses of the system to strain and pressure.

\section{density of states and Fermi surface}
\begin{figure*}[htbp]
  \centering
  \includegraphics[width=0.8\linewidth]{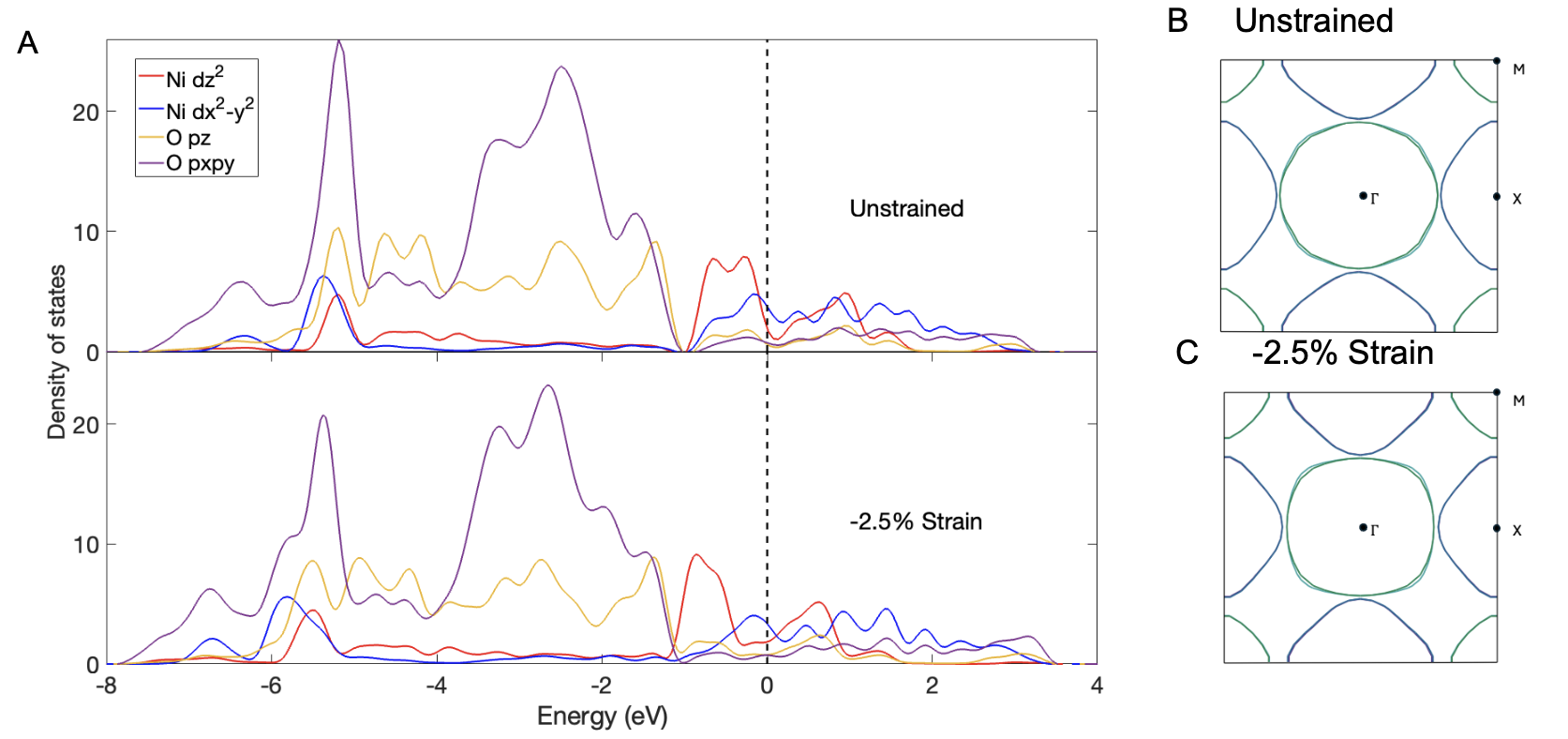}
  \caption{\large
        \textbf{Density of States and Fermi Surface of Unstrained and -2.5\% Strained $\text{La}_3\text{Ni}_2\text{O}_7$.} 
        \textbf{A:} Density of states (DOS) for unstrained and -2.5\% strained $\text{La}_3\text{Ni}_2\text{O}_7$. Under strain, the peak associated with the $d_{z^2}$ orbital shifts downward from just below the Fermi surface, as indicated by the red line. 
        \textbf{B:} Fermi surface of unstrained $\text{La}_3\text{Ni}_2\text{O}_7$. 
        \textbf{C:} Fermi surface of -2.5\% strained $\text{La}_3\text{Ni}_2\text{O}_7$. Unlike the high-pressure phase, no new hole pocket is observed around the $\Gamma$-point under compressive strain. }
  \label{Dos2}
\end{figure*}

To further investigate the orbital contributions near the Fermi level, we analyzed the density of states (DOS) for Ni \(3d_{z^2}\), \(3d_{x^2-y^2}\), and O \(2p_z\), \(2p_x\), and \(2p_y\) orbitals, as shown in Figure~\ref{Dos2}. Under a 2.5\% compressive strain, the DOS peak of Ni \(3d_{z^2}\) shifts downward relative to the Fermi level compared to the unstrained case. This shift corresponds to the downward movement of the \(\gamma\) bands, which are predominantly composed of Ni \(3d_{z^2}\) orbitals, away from the Fermi level, indicating an increased occupation of the \(d_{z^2}\) orbitals.

In terms of the Fermi surface, the -2.5\% strain does not introduce a new hole pocket around the \(\Gamma\) point, which appears under high pressure due to the upward shift of the \(\gamma\) bands. The most notable change observed is the reshaping of the Fermi surface around the \(\Gamma\) point, which is primarily associated with the Ni 3$d_{x^2-y^2}$ \(\beta\) bands.

\section{Oxygen 2$p_z$ orbital analysis}
\begin{table}[h]
\centering
\caption{\large Comparison of orbital contributions at the $\Gamma$ point for unstrained and strained -2.5\% cases. The Fermi energy is 8.29 eV for the unstrained system and 8.17 eV for the strained system. "Inner Oxygen" refers to the apical inner oxygen atoms, while "Outer Oxygen" refers to the apical outer oxygen atoms, as illustrated in Fig.~2A.}
%\begin{adjustbox}{max width=\textwidth}
\begin{tabularx}{1.08\textwidth}{XXXXX|XXXXX}
    %\end{tabularx}{ccccc|ccccc}
\toprule
\textbf{Unstrained Band} & \textbf{ Ni \(3d_{z^2}\)} & \textbf{O \(2p_z\)} & \textbf{Inner O \(2p_z\)} & \textbf{Outer O \(2p_z\)} & \textbf{Strained Band} & \textbf{ Ni \(3d_{z^2}\)} & \textbf{ O \(2p_z\)} & \textbf{Inner O \(2p_z\)} & \textbf{Outer O \(2p_z\)} \\ 
\hline
1 (7.46eV) & 0.714 & 0.094 & 0 & 0.052 & 1 (7.17eV) & 0.43 & 0.162 & 0 & 0.028 \\ 
2 (7.56eV) & 0.550 & 0.148 & 0 & 0.08 & 2 (7.28eV) & 0.546 & 0.201 & 0 & 0.092 \\ 
3 (8.10eV) & 0.628 & 0.136 & 0 & 0.128 & 3 (7.68eV) & 0.598 & 0.152 & 0 & 0.144 \\ 
4 (8.14eV) & 0.623 & 0.148 & 0 & 0.14 & 4 (7.70eV) & 0.597 & 0.163 & 0 & 0.152 \\ 
5 (8.49eV) & 0.654 & 0.205 & 0.148 & 0.012 & 5 (8.26eV) & 0.64 & 0.242 & 0.164 & 0.012 \\ 
6 (8.53eV) & 0.648 & 0.208 & 0.14 & 0.024 & 6 (8.29eV) & 0.642 & 0.238 & 0.156 & 0.02 \\ 
7 (9.28eV) & 0.498 & 0.246 & 0.136 & 0.088 & 7 (8.86eV) & 0.496 & 0.27 & 0.136 & 0.096 \\ 
8 (9.31eV) & 0.502 & 0.254 & 0.136 & 0.1 & 8 (8.87eV) & 0.498 & 0.275 & 0.136 & 0.104 \\ 
\hline
\end{tabularx}
%\end{adjustbox}
\label{tab:orbital_comparison}
\end{table}
Previous studies have shown that in La$_3$Ni$_2$O$_{7-\delta}$, oxygen deficiencies ($\delta$) exceeding 0.08 lead to an insulating state~\cite{doi:10.1143/JPSJ.64.1644}. Likewise, in thin films, the absence of an ozone annealing procedure results in insulating behavior~\cite{Ko2024,liu2025superconductivitynormalstatetransportcompressively}. Research suggests that oxygen vacancies predominantly occur at the apical inner sites, significantly suppressing superconductivity~\cite{zhang2023work2,bhatt2025resolvingstructuraloriginssuperconductivity}. These observations highlight the critical role of hybridization between Ni \(3d\) orbitals and O \(2p\) orbitals in governing both the normal-state transport properties and the underlying mechanism of superconductivity.

Since transport and superconductivity are primarily governed by the electronic bands near the Fermi level, which predominantly comprise Ni \(3d_{x^2-y^2}\) and \(3d_{z^2}\) orbitals, theoretical studies have identified the Ni \(3d_{z^2}\) orbital as the primary contributor to superconducting \(s\pm\) pairing~\cite{Luo2024,PhysRevLett.131.126001,PhysRevLett.132.106002,Lu2024,PhysRevB.109.045154}. This raises a compelling question about the role of the oxygen \(2p_z\) orbitals in these bands and how their contributions are modified under strain.To address this, we analyzed the eight bands near the Fermi level, which predominantly consist of \(3d_{z^2}\) character at the \(\Gamma\) point, and extracted the contributions of oxygen \(2p_z\) orbitals. The results, summarized in Table~\ref{tab:orbital_comparison}, reveal that in the unstrained case, the four bonding \(3d_{z^2}\) bands below the Fermi level exhibit a minimal contribution from oxygen \(2p_z\) orbitals, approximately 0.1, primarily originating from the apical outer oxygen atoms, with zero contribution from the apical inner oxygen atoms. Conversely, in the four anti-bonding \(3d_{z^2}\) bands above the Fermi level, the contribution of oxygen \(2p_z\) orbitals increases to approximately 0.2, predominantly from the apical inner oxygen atoms. When a compressive strain of \(-2.5\%\) is applied, notable changes occur in the oxygen \(2p_z\) contributions. For the two bonding bands near the Fermi level, the contribution from the apical outer oxygen \(2p_z\) orbitals increases by approximately 10\%. Similarly, for the two anti-bonding bands near the Fermi level, the contribution from the apical inner oxygen \(2p_z\) orbitals also increases by about 10\%. These observations highlight the strain-induced redistribution of orbital contributions, underscoring the significant role of apical oxygen orbitals in modulating the electronic structure near the Fermi level.

\end{document}